# Manifestation of geometric frustration on magnetic and thermodynamic properties of pyrochlores $Sm_2 X_2O_7$ (X=Ti, Zr)

Surjeet Singh[1], Surajit Saha[2], S. K. Dhar[3], R. Suryanarayanan[1], A. K. Sood[2,*] and A. Revcolevschi[1]

[1]Université Paris-Sud, Laboratoire de Physico-Chimie de l'Etat Solide, ICMMO, UMR8182, CNRS, Bât 414, 91405 Orsay, France.

[2]Department of Physics, Indian Institute of Science, Bangalore 560012, India.

[3]Tata Institue of Fundamental Research, Mumbai 400005, India.

PCAS: 75.30.-m, 75.40.Cx, 78.30.-j

We present here magnetization, specific heat and Raman studies on single-crystalline specimens of the first pyrochlore member $Sm_2Ti_2O_7$ of the rare-earth titanate series. Its analogous compound $Sm_2Zr_2O_7$ in the rare-earth zirconate series is also investigated in the polycrystalline form. The Sm spins in $Sm_2Ti_2O_7$ remain unordered down to at least T = 0.5 K. The absence of magnetic ordering is attributed to very small values of exchange ($\theta_{cw}$ ~ -0.26 K) and dipolar interaction ($\mu_{eff}$ ~ 0.15 $\mu_B$) between the $Sm^{3+}$ spins in this pyrochlore. In contrast, the pyrochlore $Sm_2Zr_2O_7$ is characterized by a relatively large value of Sm-Sm spin exchange ($\theta_{cw}$ ~ - 10 K); however, long-range ordering of the $Sm^{3+}$ spins is not established at least down to T = 0.67 K, due to frustration of the $Sm^{3+}$ spins on the pyrochlore lattice. The ground state of $Sm^{3+}$ ions in both pyrochlores is a well-isolated Kramer's doublet. The higher-lying crystal field excitations are observed in the low-frequency region of the Raman spectra of the two compounds recorded at T = 10 K. At higher temperatures, the magnetic susceptibility of $Sm_2Ti_2O_7$ shows a broad maximum at T = 140 K while that of $Sm_2Zr_2O_7$ changes monotonically. Whereas $Sm_2Ti_2O_7$ is a promising candidate for investigating spin-fluctuations on a frustrated lattice as indicated by our data, the properties of $Sm_2Zr_2O_7$ seem to conform to a conventional scenario where geometrical frustration of the spin exclude their long-range ordering.

## 1. INTRODUCTION

Geometrical frustration is presently one of the key research areas in condensed matter physics[1]. The rare-earth titanate pyrochlores $R_2Ti_2O_7$ (R = Gd to Yb) are considered to be model three-dimensional geometrically frustrated systems due to their interesting and rather unconventional magnetic properties[2]. In these insulating pyrochlores, the frustration of antiferromagnetic (for R = Gd, Tb, and Er) or ferromagnetic (R = Dy, Ho and Yb) interaction between neighboring R-spins results in various complex magnetic ground states depending on the R-ion. These include spin-liquid phase or cooperative paramagnetism[3-5] in $Tb_2Ti_2O_7$ which shows signs of spin-crystallization upon compression[6] and phase transition at $T_N \sim 3$ K as well as spin-wave excitations under high magnetic field[7], a first-order phase transition in the spin-dynamics[8] of $Yb_2Ti_2O_7$, complex antiferromagnetic order believed to be arising from the "order by disorder" mechanism[9] in $Er_2Ti_2O_7$, multiple magnetic transitions[10,11] below T = 1 K in $Gd_2Ti_2O_7$, crystal-field split non-magnetic ground state[12] in $Tm_2Ti_2O_7$ and *spin-ice* state in $Dy_2Ti_2O_7$ and $Ho_2Ti_2O_7$ titanates[13, 14] where the disorder in the $R^{3+}$ spin arrangement maps onto the problem of the proton arrangement in common water-ice, first considered by Linus Pauling[15].

While the exchange between the rare-earth spins is antiferromagnetic for almost all the titanate pyrochlores[16], the differences in the onsite anisotropy of R-ions and the presence of residual interactions, such as, dipolar interactions and/or spin fluctuations at low-temperature, cause these titanate pyrochlores to behave very differently from each other, presenting in totality a wide spectrum of interesting and rather unconventional magnetic properties. Surprisingly, the first pyrochlore member of the $R_2Ti_2O_7$ series, namely $Sm_2Ti_2O_7$ (STO), which is located on a borderline between the stable pyrochlore structure of the heavier R-ions (Gd to Lu) and the monoclinic structure of the lighter R-ions (La, Pr and Nd), has not been the subject of any detailed investigations, so far. In one of our recent works based on oriented single-crystals of the metallic pyrochlore $Sm_2Mo_2O_7$, we find evidences of an ordered spin-ice state of Sm spins at low-temperatures [17], which motivated us to undertake a parallel study in the pyrochlore STO and its analogous compound in the zirconate pyrochlore series, namely $Sm_2Zr_2O_7$ (SZO), for comparison. The pyrochlore $La_2Zr_2O_7$ (LZO) is also studied as a non-magnetic reference compound.

## 2. EXPERIMENTAL DETAILS

Polycrystalline samples of STO, SZO and LZO were prepared using the standard solid-state reaction route by heating them at temperatures near 1400 °C for 24 -36 hrs with one or two intermediate grindings. The starting oxides in each case were at least 99.9% pure commercial products from *Alfa Aesar*. Single crystals of STO were grown using the optical floating-zone technique in an infrared image furnace. Due to the high melting temperatures of LZO and SZO, single crystals of these pyrochlores could not be grown. The phase purity of the samples is determined by powder X-ray diffraction. All three compounds crystallize with the cubic pyrochlore structure. In the X-ray diffraction pattern of $La_2Zr_2O_7$, few weak intensity lines due to $ZrO_2$ are also observed. Lattice parameters of 10.227 Å, 10.593 Å and 10.796 Å, for STO, SZO and LZO, respectively, obtained from their powder x-ray diffractions are in close agreement with previously reported values for these pyrochlores[18]. In the case of STO, a single crystal piece was crushed into powder for X-ray diffraction. The magnetization measurements were carried out on a Quantum Design SQUID magnetometer between 2 to 300 K. The specific heat was measured between T = 0.5 and 300 K on a Quantum Design PPMS. Raman spectra from a (111) oriented single-crystal sample of STO and from a polycrystalline sample of SZO were recorded at T = 10 K in a back-scattering geometry using the 514.5 nm line of an argon ion laser. The sample was mounted on the cold finger of a CTI-Cryogenics Closed Cycle Refrigerator whose temperature was controlled by a CRYO-CON 34 temperature controller with Si-diode as the temperature sensor. The scattered light was analyzed using a computer controlled DILOR Raman spectrometer consisting of three holographic gratings (1800 grooves/mm) coupled to a liquid-nitrogen cooled Charged Coupled Device.

## 3. RESULTS

### (3.1) Specific Heat

Specific heat (C) data from a single crystal sample of STO and polycrystalline samples of SZO and LZO are shown in Fig. 1 as a function of temperature from T = 2 K to 300 K. The specific heat of SZO remains larger than that of non-magnetic LZO over the entire temperature range of measurements. Since the lattice specific heat of these two homologous compounds is expected to be comparable to each other, the excess specific heat in SZO is, therefore, of magnetic origin, due to Schottky-like contribution arising from a large crystal field (CF) splitting in the lowest J = 5/2 multiplet of $Sm^{3+}$. A similar excess in the specific heat of STO over that of LZO is also evident from Fig. 1. At temperatures above 150 K the specific heat of LZO seems to exceed that of STO, which may appear unrealistic at first sight; however, it is due to a smaller molecular mass of STO compared to LZO.

The specific-heat of magnetic insulators, such as STO and SZO, can be expressed as an algebraic sum of three terms: (1) the nuclear ($C_N$), (2) the lattice ($C_{latt}$) and (3) the magnetic ($C_{mag}$) terms. The $C_N$ contribution in $Sm^{3+}$ compounds is significant only below T = 0.4 K [19], therefore, the measured specific heat of STO (SZO) in the temperature range 0.5 K to 300 K is due to the lattice and magnetic contributions only. In order to determine the $C_{mag}$ contribution in the specific heats of STO and SZO, we have approximated the $C_{latt}$ contribution in these compounds by the measured specific heat of non-magnetic insulator LZO between T = 2 K and 300 K. To take into account the differences in the molecular weights of these compounds, we scaled the specific-heat of LZO by a factor $(M_{mag}/M_{LZO})^n$, where $M_{mag}$ and $M_{LZO}$ are the molecular weights of STO (or SZO) and LZO, respectively. As shown in Fig. 2, the specific heat of LZO below T = 10 K is negligibly small; consequently, the $C_{mag}$ values of STO and SZO in this temperature range are independent of the value chosen for index 'n'. In order to get a very rough estimate of $C_{mag}$ at higher temperatures, we used n = ½ in the case of SZO. We find that using n = 3/2, derivable from Debye $T^3$ law which is applicable only at very low-temperatures, resulted in large negative values of $C_{mag}$ at higher temperatures. Similarly in STO we find that n = ½ seem to work best. Since, the focus of present studies is mainly on the low-temperature behavior of Sm spins in these geometrically frustrated pyrochlores, an accurate determination of $C_{mag}$ at higher temperatures is left for future studies. The associated entropy change $\Delta S_{mag}$ in STO (SZO) in the low-temperature range [$T_0$, T] K is calculated using $\Delta S(T) = \int_{T_0}^{T} \frac{C_{mag}}{T} dT$, where $T_0$ (~ 0.5 K) is the lowest temperature in our measurements.

The temperature variation of $C_{mag}$ and $\Delta S_{mag}$ of STO and SZO, in the low temperature range, is shown in Fig. 2. At low-temperatures $C_{mag}$ of STO shows a sharp increase upon cooling below T = 2K and in the temperature range 2 K to 10 K it is almost negligible. Upon heating the sample above T = 10 K, $C_{mag}$ increases strongly and exhibits a broad anomaly near T = 60 K (Fig. 4b). The increase in $C_{mag}$ below T = 2 K is presumably due to short-range Sm-Sm spin correlations; however, no sign of Sm spins ordering or freezing is present down to our lowest measurement temperature of T = 0.5 K. The change in the spin or magnetic entropy ($\Delta S_{mag}$) of STO between T = 0.5 K and 10 K is only 13 % of Rln2 (Fig. 2b).

The $C_{mag}$ behavior in SZO is qualitatively similar to that in STO. In the low-temperature range $C_{mag}$ is dominated by short-range magnetic correlations of the Sm-spins which are responsible for the observed increase in $C_{mag}$ upon cooling below T = 10 K.

However, no sign of long-range ordering or freezing of the Sm spins could be seen down to the lowest measured temperature of T = 0.67. The change in the spin entropy ($\Delta S_{mag}$) of SZO between T = 0.67 K and 10 K is 3 J(Sm-mol)$^{-1}$K$^{-1}$. In the high-temperature range $C_{mag}$ of SZO exhibits a broad anomaly near T = 100 K (Fig. 4a). This anomaly, and a similar anomaly in $C_{mag}$ of STO near T = 60 K, is arising from the thermal occupation of higher lying CF levels of Sm$^{3+}$. To investigate the CF scheme in these Sm pyrochlores, we carried out Raman scattering studies, which is a powerful technique of probing CF excitations in solids [20].

**(3.2) Raman Scattering Results and Schottky Heat Capacity Anomaly**

Raman spectrum of a single crystalline sample of STO (incident laser beam parallel to the [111] crystallographic axis) and a polycrystalline sample of SZO, recorded at T = 10 K, are shown in Fig. 3. Raman spectrum due to STO is characterized by the presence of *six* Raman-active phonon modes labeled $p_i$ (*i* =1 to 6) in agreement with the factor group analysis of the Raman-active phonon modes for the pyrochlore lattice [21]. The assignment of various Raman-active phonon modes is done after ref. [21-23] (Table 1). In addition to the *six* Raman-active phonon modes, *four* well-resolved modes near $\Delta\omega$ = 87, 132, 158 and 270 cm$^{-1}$, labeled as $c_i$ (*i* = 1 to 4), due to CF excitation of Sm$^{3+}$ are also observed. A similar observation is made in the Raman spectrum of SZO recorded at T = 10 K, which shows the presence of *six* Raman-active phonon modes, in good agreement with the T = 300 K spectrum reported in ref.[24], and *four* CF modes, not reported so far, near 85, 130, 165 and 210 cm$^{-1}$ (these modes are labeled as $e_i$ (*i* = 1 to 4)). The phonon modes in SZO are relatively broad, even at a low temperature of T = 10 K, which can be attributed partly to the polycrystalline nature of the SZO sample and partly also to the increasing tendency, in the rare-earth zirconates (R$_2$Zr$_2$O$_7$) series, to evolve towards the defect-fluorite structure with decreasing R$^{3+}$ ionic radius [18, 24].

The compounds of stoichiometry $R_2X_2O_7$ crystallize with the pyrochlore structure if the ratio of ionic radii of the cations $R^{3+}$ and $X^{4+}$ lies within the "pyrochlore stability-field"[18], defined as $1.46 < (\xi = r^{3+}/r^{4+}) < 1.80$. When $\xi$ is close to 1.46, the defect-fluorite structure is favored over the pyrochlore structure, as in the heavier rare-earth members (R = Tb-Lu) of the $R_2Zr_2O_7$ series of compounds. In the defect-fluorite structure ($Fm3m$), the anions (cations) statistically occupy the 8c (4a) sites in the cubic sublattice of $F^{2-}$($Ca^{4+}$) of the $CaF_2$-type structure. On the other hand, in the pyrochlore structure ($Fd\overline{3}m$), the cations $R^{3+}$ and $X^{4+}$ selectively occupy the 16d and 16c sites, albeit, with some degree of anti-site disorder. Further, the anionic sublattice shows large deviation from its cubic symmetry in the $CaF_2$-type structure. The deviation arises due to the movement of six $O^{2-}$ ions in the $R_2X_2O_7$ formula unit to a new position 48f of site symmetry lower than cubic. The seventh anion in each formula unit occupies the 8b site, leaving the 8a sites in the pyrochlore structure systematically vacant.

STO is a borderline pyrochlore compound because of its large $\xi$ value ($\xi$(STO) $\approx$ 1.78); with the following consequences: (1) the antisite disorder in STO is *minimal* due to a huge disparity between the ionic radii of $Sm^{3+}$ and $Ti^{4+}$, and (2) the distortion of its anionic sublattice is larger than for any other member of the titanate pyrochlore series. The coordination polyhedron of $Sm^{3+}$ in STO consists of *six* O(48f) anions that form a puckered hexagonal crown and *two* O(8b) anions that form a linear O(8b)-Sm-O(8b) bond parallel to the <111> crystallographic axis, which is perpendicular to the mean plane of the O(48f) crown. Since the bond length $d_{Sm-O(8b)}$ is about 15 % shorter then $d_{Sm-O(48f)}$, the resulting crystal electric field at the $R^{3+}$ site is trigonal ($D_{3d}$) rather than cubic which splits the (2J + 1)-fold degenerate J-multiplets in a free $Sm^{3+}$ ion into groups of Kramer's doublets.

The magnitude of the overall CF splitting of the J-multiplets in the titanate pyrochlores is typically of the order of few hundred wavenumber. For example, in the pyrochlore

Ho$_2$Ti$_2$O$_7$, the first CF excitation in the lowest J = 7 multiplet of Ho$^{3+}$ is near 160 cm$^{-1}$ and magnitude of the overall CF splitting is about 650 cm$^{-1}$ ($\approx 10^3$ K) [25].

The number of observed CF modes in the Raman Spectra of both of STO and SZO exceeds the number of CF excitations in the lowest J = 5/2 multiplet of Sm$^{3+}$, which indicates that some of the weak CF modes in the spectra of these compounds are arising from electronic transitions between the excited CF levels of the J = 5/2 multiplet and low-lying CF levels of the J = 7/2 multiplet. The CF split low lying levels of J = 7/2 multiplet may not be very far from the top of CF split J = 5/2 multiplet because of the small inter-multiplet width of Sm$^{3+}$ and large CF splittings, typical of these pyrochlores. Thus, the assignment of the various CF modes to corresponding electronic transitions in the various J-multiplets of Sm$^{3+}$ in both STO and SZO may be complicated depending on the selection rules and we leave it for a future work.

In Fig. 4 we have shown the magnetic specific heat of STO and SZO over the entire temperature range up to T = 300 K. The broad anomaly in $C_{mag}$ of STO near T = 60 K and in SZO near T = 100 K is a Schottky-like anomaly arising due to thermal (de)population of excites CF levels of Sm$^{3+}$. We calculated a multi-level Schottky anomaly to compare with the experimental $C_{mag}$ in Fig. 4. We find that the position and height of a Schottky anomaly calculated by taking the $e_2$ = 130 cm$^{-1}$ and $e_4$ = 210 cm$^{-1}$ modes in the Raman spectrum of SZO as the positions of first and second excited CF levels of Sm$^{3+}$ matches closely with the anomaly in the experimental $C_{mag}$ as shown in Fig. 4a. The deviation in the low temperature side of the anomaly, where our estimate of $C_{mag}$ is reasonably good, is due to short-range Sm$^{3+}$ spins correlations discussed above. In STO, on the other hand, where the Schottky peak appears at a much lower temperature of 60 K, a good matching with the experimental $C_{mag}$ is obtained by taking the first excited CF level near 50 cm$^{-1}$ and the second excited CF level near $c_2$ = 132 cm$^{-1}$ (Fig. 4b). It is, therefore, possible that a CF mode corresponding to the first CF level of $Sm^{3+}$ in STO lies near or below 50 $cm^{-1}$. Large deviations on the high temperature side of the anomaly are presumably due to errors associated with the exact determination of $C_{mag}$ in this temperature range, as discussed earlier. Unlike SZO, however, the overlap of the calculated Schottky curve and experimental $C_{mag}$ of STO in the low temperature side of the anomaly is reasonably good which is expected as the short-range Sm spins' correlations in STO appears only below T = 2 K.

Based on our specific heat data we can draw the following preliminary conclusions: (1) in both, STO and SZO, the Kramer's doublet ground state of $Sm^{3+}$ is well-isolated from the higher lying doublets, (2) in STO the short-range magnetic correlations between the $Sm^{3+}$ spins develop below T = 2 K, (3) in SZO, on the other hand, $C_{mag}$ is appreciably large, even at T = 10 K. Since the position of Schottky anomaly in SZO is even higher than that in STO, one cannot attribute the large $C_{mag}$ at low-temperatures to thermal occupation of higher lying CF levels. Clearly, the short-range magnetic correlations in SZO are significant even at higher temperatures, (4) no long-range magnetic ordering of Sm spins is seen down to T = 0.67 K, which is an indication of a rather strong geometrical frustration of $Sm^{3+}$ spins at play in this pyrochlore. We investigated the magnetic susceptibility of these pyrochlores to get a mean-field estimate the Sm-Sm spin exchange in the two compounds. We shall see in the next section that our estimate of Sm-Sm spin exchange in these compounds is in a good agreement with the temperature variation of low-temperature specific heat in these pyrochlores.

**(3.3) Magnetic Susceptibility**

The DC magnetic susceptibilities ($\chi$) of a single crystalline sample of STO (open circles) and of polycrystalline SZO (open squares) are shown in Fig. 5 as a function of temperature. A room-temperature susceptibility of roughly $10^{-3}$ emu/Sm-mol, for both STO and SZO, is expected of a +3 oxidation state of the Sm-ions in both compounds, as implied from the charge neutrality of their pyrochlore formula units. $\chi$ of SZO exhibits a monotonic increase upon cooling below T = 300 K down to our lowest measurement temperature of 2 K. In contrast, $\chi$ of STO exhibits a rather non-trivial temperature dependence dominated by the presence of a broad maximum centered at T = $T_{max}$ = 140 K (inset of Fig. 5). Upon cooling below $T_{max}$, $\chi$ decreases by nearly 12 % of its value at $T_{max}$, followed by a steep rise below T = 50 K.

In order to understand the origin of the susceptibility maximum in STO we studied a highly diluted STO samples, namely $Sm_{0.05}Y_{1.95}Ti_2O_7$ (SYTO). In this SYTO sample, nearly 98.5 % of the $Sm^{3+}$ ions are replaced by the non-magnetic $Y^{3+}$ ions. If we assume that 2.5 % of $Sm^{3+}$ ions in SYTO are well-isolated from each other, then all the magnetic characteristics of SYTO can be attributed to the CF derived single-ion properties of the Sm-spins in the pyrochlore structure. The susceptibility per $Sm^{3+}$ ion of the SYTO sample is shown in the inset of Fig. 5. For comparison, the susceptibility of the STO sample is also plotted on the same scale. A small hump in $\chi$ of the SYTO sample, at roughly the same temperature as the maximum in the susceptibility of STO, is clearly decipherable, which may indicate that the $\chi$-maximum in STO is related to the CF effects. Since clustering of the Sm ions in the SYTO cannot be completely ruled out, similar studies on further diluted samples are needed to confirm this assertion.

The temperature dependence of $\chi$ in SZO is found to be non-Curie-Weiss-like, which is typical of $Sm^{3+}$ compounds due to a narrow J-multiplet width of $Sm^{3+}$. The susceptibility of SZO is therefore fitted to an expression of the form:

$\chi = \chi_{cw} + \chi_{vv,}$ (1),

where $\chi_{cw} = N_A\mu_{eff}^2/3k_B(T - \theta_P)$ is the Curie-Weiss contribution arising from the Zeeman-split lowest J = 5/2-multiple of $Sm^{3+}$ and $\chi_{vv} = 20\mu_B^2/7k_B\Delta_J$ is the temperature-independent van Vleck contribution from the first excited J = 7/2-multiplet of $Sm^{3+}$ at $\Delta_J \approx 4T_{room}$ ($T_{room} \sim 300$

K). In these expressions $N_A$, $k_B$ and $\mu_B$ denote Avogadro's number, Boltzmann constant and Bohr magneton, respectively. The effective $Sm^{3+}$ magnetic moment ($\mu_{eff}$), Curie-Weiss temperature ($\theta_P$) and the inter-multiplet splitting width ($\Delta_J$) are the fitting parameters. Using such a fitting procedure in the temperature range between 50 K and 300 K, we obtained a value of 0.50 $\mu_B$ for $\mu_{eff}$, and -30 K and 1390 K for $\theta_P$ and $\Delta_J$, respectively. While the value of $\Delta_J$ is comparable to the theoretical value, the $\mu_{eff}$ value is somewhat smaller than a free-ion value of 0.83 $\mu_B/Sm^{3+}$. The negative value of $\theta_P$ indicates an antiferromagnetic exchange between the Sm spins, which is in accordance with the linear variation of magnetization as a function of applied field, shown in Fig. 6. The value of $\theta_p$ was found to depend on the choice of the lowest temperature $T_0$ of the fitting interval [$T_0$ – 300 K], varying from $\theta_p$ = -30 K for $T_0$ = 50 K to $\theta_p$ = -45 K for $T_0$ = 100 K. Both, the dependency of $\theta_P$ on the fitting temperature-range and a reduced value of the effective $Sm^{3+}$ moment, are probably related to the large CF splitting of the lowest J= 5/2 multiplet, which is not taken into account in our fitting procedure. Even so we could not determine the exact value of $\theta_P$, the least that we can conclude from our analysis is that the magnitude of $\theta_P$ in SZO is roughly of the order of 10 K or so, consistent with our observation of short-range Sm spins correlations in the low temperature specific heat data.

Since, the magnetic correlations in STO develop well below T = 5 K, we fitted the low-temperature $\chi$ data of STO using expression (1) in the temperature range [5, 20] K. In this case, $\chi_{cw}$ in expression (1) is the Curie-Weiss susceptibility arising mainly from the well-isolated Kramer's doublet ground state of $Sm^{3+}$, the excited CF levels being much higher in temperature than our fitting temperature-range. Using such a fitting procedure we obtained $\mu_{eff}$ = 0.15 $\mu_B$, $\chi_{vv}$ = 9.4e-4 emu/mol and $\theta_P$ = -0.26 K. The value of the effective $Sm^{3+}$ moment in the Kramer's doublet ground state of STO is significantly smaller than the free-ion value of 0.83 $\mu_B$/Sm. A small negative $\theta_P$ in STO indicates a weak antiferromagnetic exchange between the Sm spins which is in accordance with the M-H behavior shown in Fig. 6.

A value of $\theta_P$ = -0.26 K in STO is smaller but comparable to $\theta_P \approx$ -1 K for the spin-ice[26] $Ho_2Ti_2O_7$. However, what makes STO unique amongst the other pyrochlore members of the $R_2Ti_2O_7$ series is the small value of $\mu_{eff}$ = 0.15 $\mu_B$/Sm, which is nearly two orders of magnitude smaller than the effective moment per $Ho^{3+}$ in $Ho_2Ti_2O_7$ in the spin-ice state. Consequently, the strength of dipolar interaction between the neighboring Sm spins in STO is vanishingly small.

**(3.4) DISCUSSION OF THE LOW TEMPERATURE PROPERTIES**

The low-temperature specific heat and magnetic susceptibility of STO suggest that short-range Sm spin correlations in this pyrochlore begin to develop roughly below T = 2 K. The absence of Sm spins' ordering or freezing at least down to T = 0.5 K is consistent with a small value of exchange ($\theta_{cw}$ = - 0.26 K) and dipolar interaction ($\mu_{eff}$ = 0.15 $\mu_B$). The spin entropy change of 0.76 J (Sm mol)$^{-1}$K$^{-1}$ in STO, between T = 0.5 and 10 K, is nearly 87 % short of the expected Rln2 value. One may argue that the low-temperature upturn in $C_{mag}$ is a precursor to the long-range ordering of the Sm spins below T = 0.5 K, which may account for the "missing" entropy in STO. In conventional magnetic systems such a presumption can be taken at its face value but, in compounds of pyrochlore structure, some caution needs to be exercised because the spin-frustration in pyrochlores can result in highly degenerate ground states characterized by their finite "zero-point" entropies. For example, the spin-ice ground state of the pyrochlores $Dy_2Ti_2O_7$ and $Ho_2Ti_2O_7$ is massively degenerate with a "zero-point" entropy of $(^1/_2)Rln(^3/_2)$. Motivated by the AC susceptibility studies in the spin-ices $Dy_2Ti_2O_7$ and $Ho_2Ti_2O_7$, which show frequency dependent features above T = 2 K, even though the strength of exchange force in these compounds is of the order of 1 K, we measured the AC susceptibility of STO between T = 2 K and 20 K at several different frequencies. The AC-$\chi$ response of STO shows absolutely no frequency dependence down to a temperature of 2 K, which means that even at T = 2 K the Sm spins are largely uncorrelated and no signs of a possible spin-freezing at lower temperatures could be detected, which is consistent with a preliminary µSR study that showed no change in the spin-relaxation-rate down to T = 2 K [26]. Since the strength of exchange and dipolar interactions in the pyrochlore STO are very weak in nature, we believe that spin-fluctuations at low-temperatures should play a key role in STO in lifting the ground state degeneracy and hence accounting for the "missing" entropy. Indeed, our preliminary specific-heat data in a dilution refrigerator revealed a power-law behavior over a broad temperature range [27]. Further low-temperature specific heat and µSR studies are in progress to address these questions.

The low temperature specific-heat and magnetic susceptibility of the pyrochlore SZO show the presence of short-range Sm spin correlations up to as high a temperature as T = 50 K; however, no magnetic ordering of the Sm spins is seen down to a temperature of 0.67 K. The geometrical frustration of the Sm spins on the pyrochlore lattice is presumably responsible for the suppression of magnetic ordering in this compound. A comparison with the pyrochlores $Nd_2Zr_2O_7$ (NZO) [28] and $Gd_2Zr_2O_7$ [29], both of which show a transition into a long-range ordered magnetic state below T = 1 K, suggest that the $Sm^{3+}$ spins in SZO are on the verge of a long-range ordering near T = 0.65 K. The change in the spin entropy ($\Delta S_{mag}$) of SZO between T = 0.67 K and 10 K is 3 J(Sm-mol)$^{-1}$K$^{-1}$, which is nearly 52 % of the expected Rln2 entropy in the Kramers doublet ground state of $Sm^{3+}$. The "missing" magnetic entropy in SZO is probably buried below T = 0.67 K as in the case of NZO where a full Rln2 entropy is recovered in the temperature range [0.1, 10] K.

## 4. SUMMARY AND CONCLUSIONS

We have studied magnetization, specific heat and Raman scattering properties of single crystalline samples of the pyrochlore STO and polycrystalline SZO. A brief summary of our results on STO is as follows: (1) Sm spins in this pyrochlore remain unordered down to at least T = 0.5 K, (2) the ground state of the $Sm^{3+}$ ion in STO is a well-isolated Kramer's doublet, (3) a value of -0.26 K for the Weiss constant ($\theta_{cw}$) and of 10 mK for the dipolar interactions in the Kramer's doublet ground state of $Sm^{3+}$ are obtained from a Curie-Weiss analysis of the low-temperature susceptibility data, (4) the strength of the exchange and the dipolar interactions are very small, which is attributed to the absence of Sm spins ordering down to at least T = 0.5 K, (5) magnetic susceptibility at higher temperatures is dominated by a maximum near 140 K, which is related to the crystal field properties of the $Sm^{3+}$ ions in the pyrochlore structure of STO, (6) the Raman spectrum recorded at T = 10 K showed the presence of crystal field modes at: 87, 132, 158 and 270 $cm^{-1}$.

A parallel study conducted on the pyrochlore SZO revealed very different low-temperature properties compared to STO. The magnetic susceptibility of this compound varies in a manner typical of trivalent Sm compounds. Both the specific heat and magnetic susceptibility data in SZO indicate presence of short-range Sm spins' correlations up to a temperature of few tens of Kelvins. However, no long-range ordering of the Sm spins is established down to T = 0.67 K, indicative of a strong geometrical frustration of the Sm-spins at play in this pyrochlore. The Raman spectrum of SZO also showed the presence of crystal field excitations of $Sm^{3+}$ at 85, 130, 164 and 210 $cm^{-1}$.

In conclusion, the pyrochlore STO presents a unique behavior due to a significant lowering of both the exchange and dipolar energy scales over the other pyrochlore members of rare-earth titanate series. These attributes of STO make it a promising candidate for investigating spin-fluctuations on a frustrated lattice. Indeed, our preliminary specific heat study in a dilution refrigerator shows a power-law behavior presumably due to spin-fluctuations of Sm spins [27]. Further detailed studies in this compound down to much lower temperatures using µSR probe and specific heat measurements should prove useful. The low-temperature properties of the pyrochlore SZO conform to a conventional scenario where geometrical frustration of the spins exclude their long-range ordering which are interesting in their own rights.

**5. ACKNOWLEDGEMENTS**

We thank the Indo-French Centre for Promotion of Advanced Research (IFCPAR) - Centre Franco-Indien pour la Promtion de la Recherche Avancée (CEFIPRA) under project no. 3108-1. AKS also thanks the Department of Science and Technology (DST), India, for financial support.

\* Corresponding author: email: asood@physics.iisc.ernet.in, Phone: +91-80-22932964

Table-1: Raman modes frequencies (in cm$^{-1}$) of $Sm_2Ti_2O_7$ and $Sm_2Zr_2O_7$

| Modes | $Sm_2Ti_2O_7$ | $Sm_2Zr_2O_7$ |
|---|---|---|
| $F_{2g}$ | 203 | 185 |
| $F_{2g}$ | 313 | 303 |
| $E_g$ | 328 | 335 |
| $A_{1g}$ | 509 | 400 |
| $F_{2g}$ | 509 | 525 |
| $F_{2g}$ | 540 | 604 |

**FIGURE CAPTIONS**

**FIG. 1** (Color online) Specific heat (C) of pyrochlores $Sm_2Ti_2O_7$, $Sm_2Zr_2O_7$ and $La_2Zr_2O_7$ is shown as a function of temperature between T = 2 K to 300 K.

**FIG. 2** (Color online) **(a)** The temperature (T) variation of specific heat (C) and magnetic specific heat ($C_{mag}$) of pyrochlores $Sm_2Ti_2O_7$ and $Sm_2Zr_2O_7$ in the low-temperature range between T = 2 K and 30 K. Specific heat of the non-magnetic pyrochlore $La_2Zr_2O_7$ is shown by a dashed curve. **(b)** The change in magnetic entropy ΔS (right axis) and $C_{mag}/T$ (left axis) of pyrochlore $Sm_2Ti_2O_7$ and $Sm_2Zr_2O_7$ shown as a function of temperature.

**FIG. 3** (Color online) Raman spectra of the pyrochlores **(a)** $Sm_2Ti_2O_7$ and **(b)** $Sm_2Zr_2O_7$ recorded at T = 10 K.

**FIG. 4** (Color online) The temperature variation of magnetic specific heat $C_{mag}$ of pyrochlores **(a)** $Sm_2Zr_2O_7$ and **(b)** $Sm_2Ti_2O_7$ in the temperature range between T = 2 K and 300 K. Dashed line in both panels shows the calculated Schottky heat capacity (see text for details).

**FIG. 5** (Color online) Susceptibility (χ) of pyrochlores $Sm_2Ti_2O_7$ and $Sm_2Zr_2O_7$ shown as a function of temperature between T = 2 K and 300 K. Inset shows a blown-up view of the susceptibility of pyrochlores $Sm_{0.5}Y_{1.95}Ti_2O_7$ and $Sm_2Ti_2O_7$ as a function of temperature between T = 2 K and 300 K.

**FIG. 6** (Color online) Isothermal magnetization (M) of the pyrochlore $Sm_2Ti_2O_7$ at T = 2 K, 70 K and 175 K and the pyrochlore $Sm_2Zr_2O_7$ at T = 2 K plotted as a function of applied magnetic field (H).

**FIG. 1**

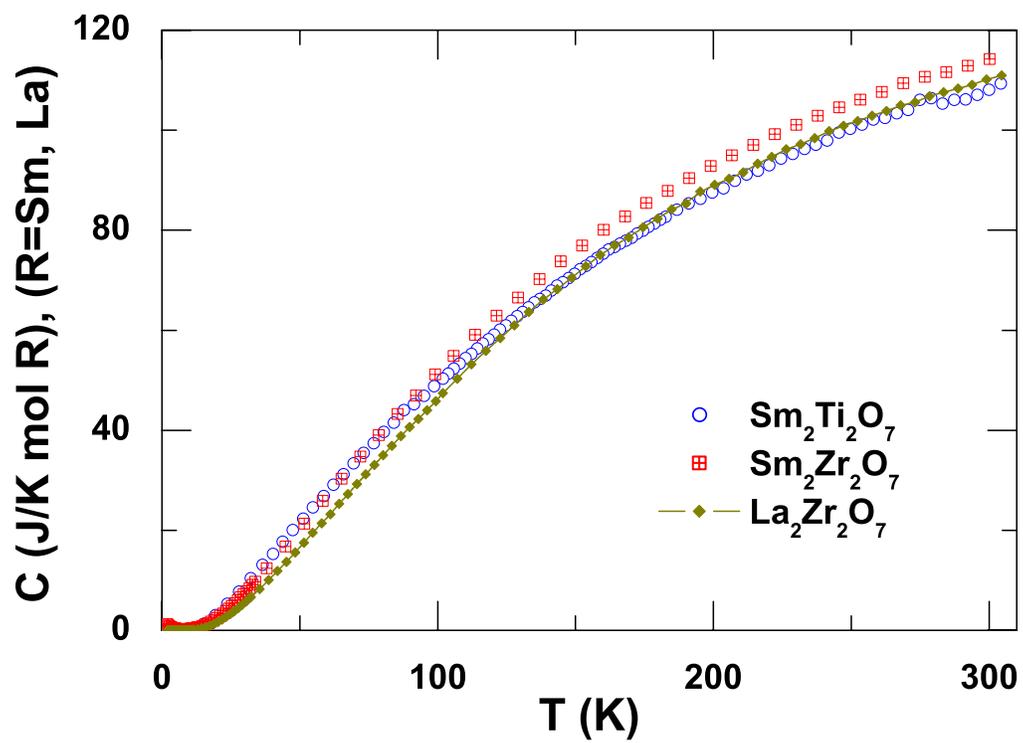

**FIG. 2**

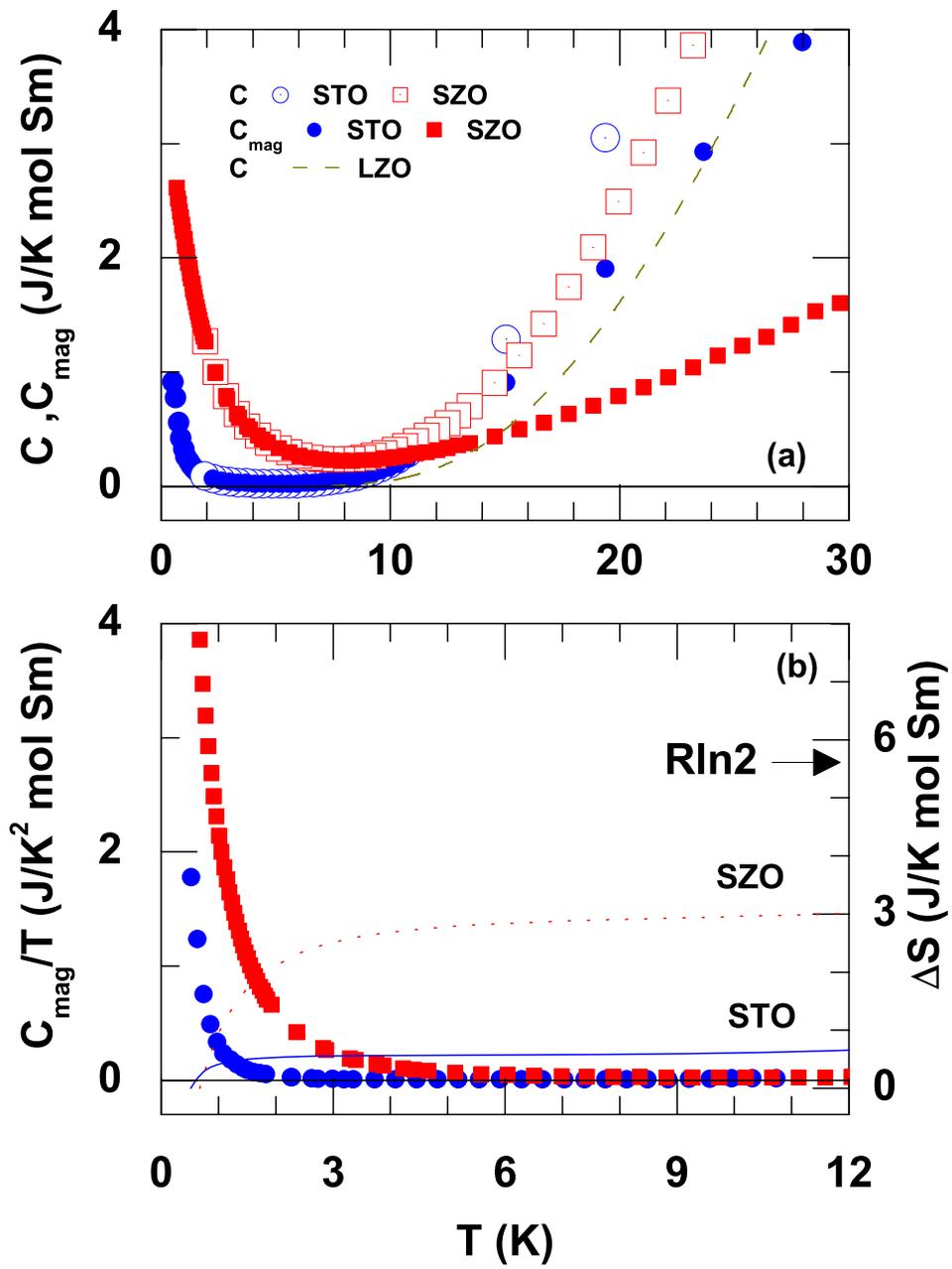

**FIG. 3**

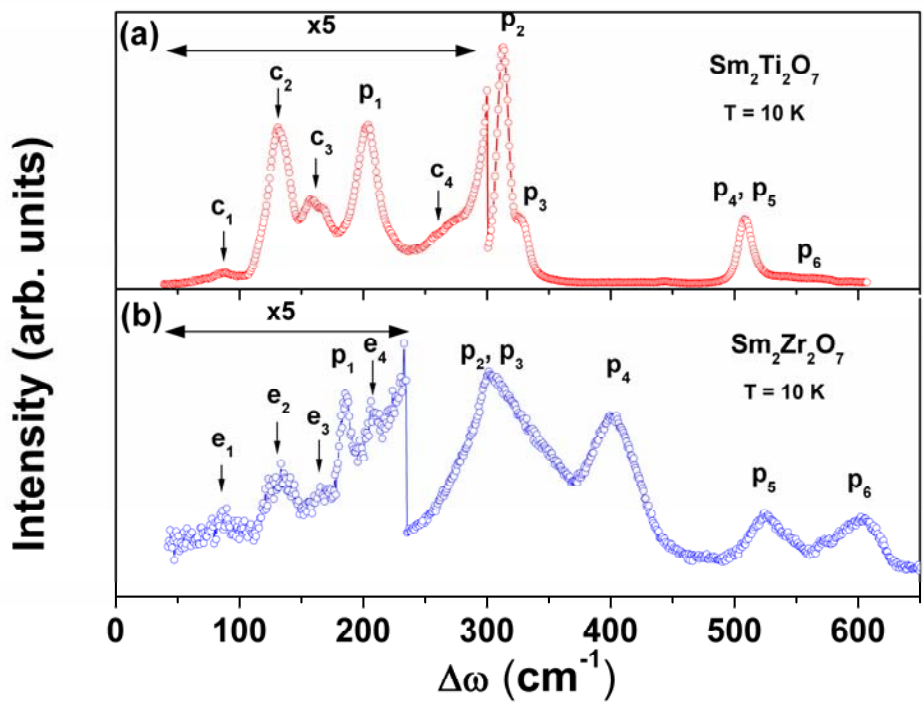

**FIG. 4**

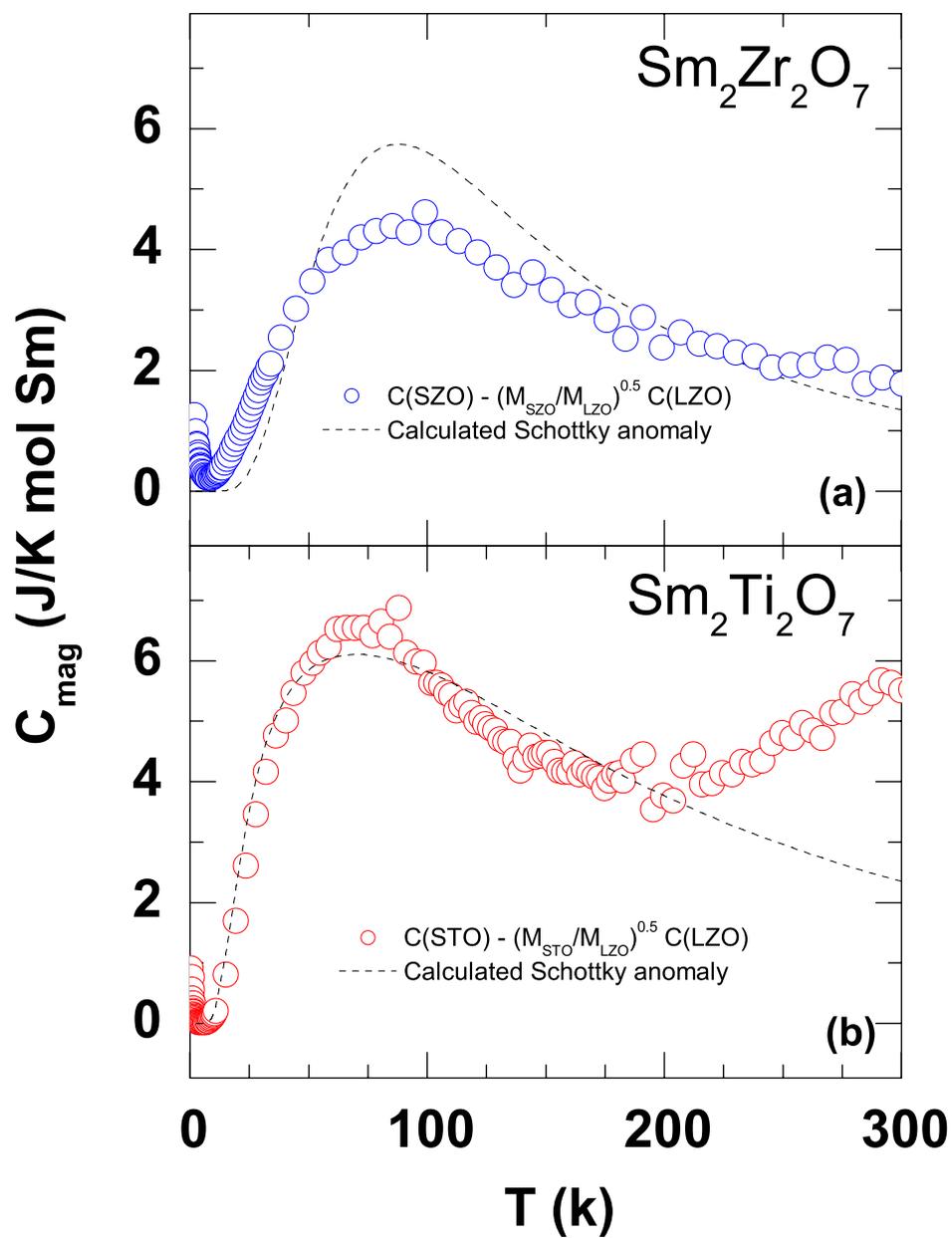

**FIG. 5**

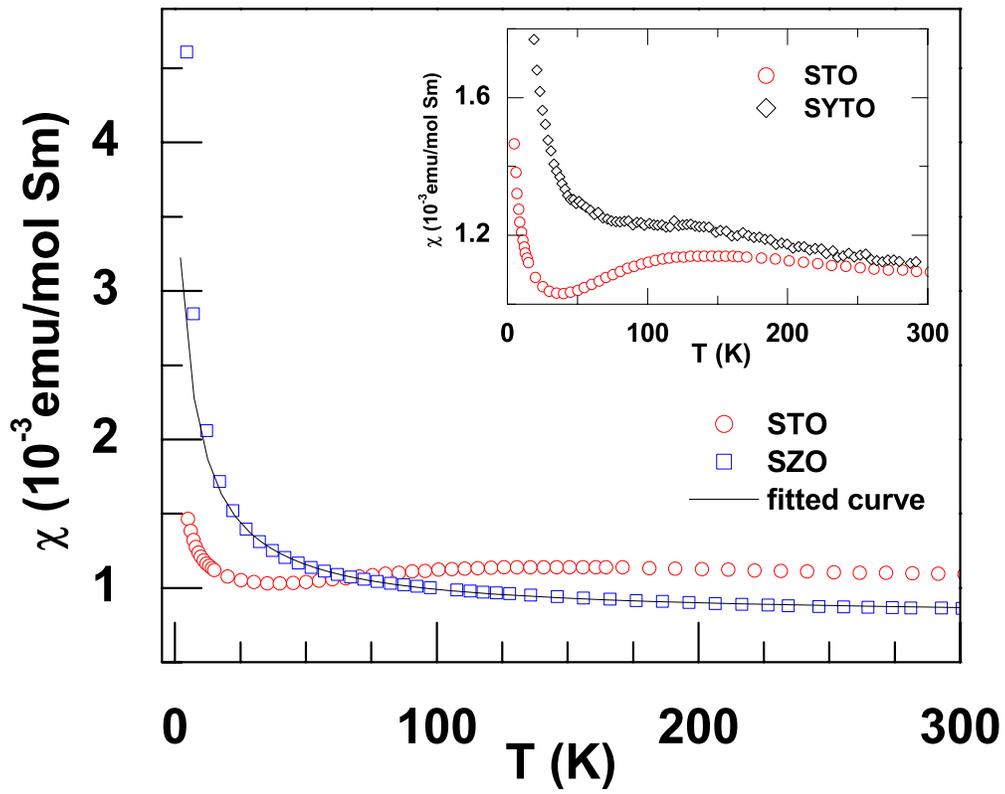

**FIG. 6**

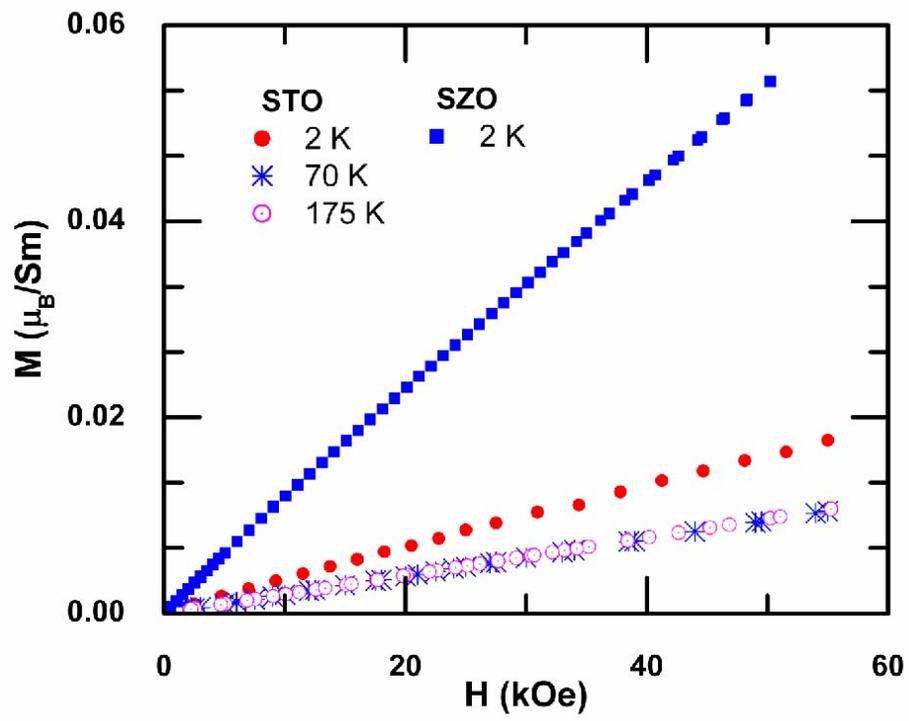